\documentclass[fleqn,usenatbib]{mnras}

\usepackage{newtxtext,newtxmath}

\usepackage[T1]{fontenc}
\usepackage{ae,aecompl}

\usepackage{graphicx}	
\usepackage{amsmath}	
\usepackage{mhchem}     

\title[Radiolysis of C$_2$H$_{4}$O$_{2}$ and DME in Cold Dark Clouds]{The Role of Radiolysis in the Modelling of C$_{2}$H$_{4}$O$_{2}$ Isomers and Dimethyl Ether in Cold Dark Clouds}

\author[A. Paulive, C. N. Shingledecker, E. Herbst]{
Alec Paulive,$^{1}$\thanks{E-mail:ap4kz@virginia.edu}
{Christopher N. Shingledecker,$^{2,3,4}$}
{Eric Herbst,$^{1,5}$}
\\
$^{1}$Department of Chemistry, University of Virginia, McCormick Road, Charlottesville, VA 22904, USA \\
$^{2}$ Max-Planck-Institute fuer Extraterrestrische Physik,
D-85748 Garching, Germany \\
$^{3}$ Institute for Theoretical Chemistry, University of Stuttgart, Pfaffenwaldring 55, 70569, Germany \\
$^{4}$ Department of Physics \& Astronomy, Benedictine College, Atchison, KS 66002, USA \\
$^{5}$ Department of Astronomy, University of Virginia, McCormick Road,  Charlottesville, VA 22904, USA \\
}

\date{Accepted XXX. Received YYY; in original form ZZZ}

\pubyear{2020}

\begin{document}
\label{firstpage}
\pagerange{\pageref{firstpage}--\pageref{lastpage}}
\maketitle

\begin{abstract}
Complex organic molecules (COMs) have been detected in a variety of interstellar sources. The abundances of these COMs in warming sources can be explained by syntheses linked to increasing temperatures and densities, allowing quasi-thermal chemical reactions to occur rapidly enough to produce observable amounts of COMs, both in the gas phase, and upon dust grain ice mantles.  The COMs produced on grains then become gaseous as the temperature increases sufficiently to allow their thermal desorption. The recent observation of gaseous COMs in cold sources has not been fully explained by these gas-phase and dust grain production routes.
Radiolysis chemistry is a possible non-thermal method of producing COMs in cold dark clouds. This new method greatly increases the modeled abundance of selected COMs upon the ice surface and within the ice mantle due to excitation and ionization events from cosmic ray bombardment. We examine the effect of radiolysis on  three C$_{2}$H$_{4}$O$_{2}$ isomers -- methyl formate (HCOOCH$_3$), glycolaldehyde (HCOCH$_2$OH), and acetic acid  (CH$_3$COOH) -- 
and a chemically similar molecule, dimethyl ether (CH$_3$OCH$_3$), in cold dark clouds. We then compare our modelled gaseous abundances with observed abundances in TMC-1, L1689B, and B1-b. 
\end{abstract}




\section{Introduction}

The isomers of C$_{2}$H$_{4}$O$_{2}$ -- methyl formate (HCOOCH$_3$), glycolaldehyde (HCOCH$_2$OH), and acetic acid (CH$_3$COOH) -- are all important chemical components of the interstellar medium. First detected by \citet{brown_discovery_1975} toward Sgr B2N, methyl formate, the most abundant of the three commonly occurring C$_{2}$H$_{4}$O$_{2}$ isomers, has been detected toward warm sources, such as Sgr B2N and Orion A, and cold molecular clouds, such as TMC-1 and L1544 \citep{brown_discovery_1975,ellder_methyl_1980,jimenez-serra_spatial_2016,soma_complex_2018}.  \citet{soma_complex_2018} determined the column density of methyl formate toward TMC-1 to be $(1.5 \pm 0.1) \times 10^{12}$ cm$^{-2}$. This column density, along with the estimated molecular hydrogen column density of $10^{22}$ cm$^{-2}$ yields a fractional abundance of  approximately $1.5 \times 10^{-10}$. Abundances of methyl formate toward L1689B and B1-b have been reported as $7.4 \times 10^{-10}$, and $2.0 \times 10^{-11}$, respectively \citep{bacmann_detection_2012,cernicharo_discovery_2012}.

As far as we know, both acetic acid and glycolaldehyde have yet to be detected toward cold molecular clouds, although  both have been observed toward Sgr B2N \citep{remijan_acetic_2002,remijan_survey_2003, remijan_survey_2004}, among other warm molecular clouds and star-forming regions. Indeed, glycolaldehyde was first detected toward Sgr B2N by \citet{hollis_interstellar_2000},  while acetaldehyde was first detected in Sgr B2, Orion KL, and W51  by \citet{mehringer_detection_1997}. 
Glycolaldehyde is a sugar-related molecule \citep{carroll_submillimeter_2010} that might serve as a precursor to the kinds of true sugars, such as ribose and deoxyribose, that are critical constituent molecules in RNA and DNA, respectively.  Acetic acid is a simple carboxylic acid, with a common functional group in biologically relevant compounds, and an amine group distant from the simplest amino acid, glycine. Another molecule that we will consider in the paper is dimethyl ether. While  not an isomer of C$_{2}$H$_{4}$O$_{2}$, it has been shown to share spatial abundance with methyl formate in areas of the ISM \citep{taquet_chemical_2017, soma_complex_2018}. 

The synthesis of COMs in warming sources in the interstellar medium is generally well understood, principally by efficient diffusion processes of radicals produced on grain surfaces by photodissociation, but also by gas-phase reactions.  The surface radicals combine to form COMs and other species, which, at still higher temperatures,  desorb thermally into the gas where they are detected. The granular synthesis is not thermally active, however, at temperatures below 20 K, since radicals - with the exception of light species such as atomic hydrogen - cannot diffuse rapidly enough to react.   A number of other approaches to the synthesis of COMs in cold regions have been reported in the literature.  These include both gas-phase chemistry, grain surface chemistry, and combinations of the two  \citep{chang_unified_2016,vasyunin_unified_2013,balucani_formation_2015}. The grain surface routes require some non-thermal mechanism to desorb molecules at low temperatures.  The major desorption processes considered include reactive desorption, photo-desorption, and cosmic ray-induced desorption via heating and sputtering, most of which are only understood at a semi-quantitative level. \citep{garrod_non-thermal_2007,hasegawa_new_1993,oberg_photodesorption_2009, kalvans_chemical_2019}.

There are a variety of low-temperature methods for producing COMs on grains prior to non-thermal desorption into the gas. While diffusive (Langmuir-Hinshelewood) grain reactions at low temperatures tend to be inefficient unless one of the reactants is an atom or diatomic molecule, non-diffusive processes do exist  \citep{hasegawa_new_1993}.  One such mechanism is the Eley-Rideal mechanism, in which species from the gas phase directly react with surface molecules  \citep{ruaud_modelling_2015}.  This process can be efficient if there are sufficient reactive species on the grain surface and if the reactions do not possess activation energy.  Another non-diffusive mechanism is the hot atom  mechanism, in which energized  granular species with enough energy to overcome diffusion barriers travel swiftly around sites on the surface until all excess energy is lost to the surface itself, or they react.

In addition to their effect on desorption, cosmic rays can induce a non-thermal chemistry as they bombard and travel through grain mantles, initially producing reactive secondary electrons. The non-thermal chemistry, known as radiolysis, has been studied experimentally by a number of groups, who utilize either energetic protons or electrons to bombard low temperature ices and produce many molecules including COMs   \citep{abplanalp_study_2016, rothard_modification_2017,boyer_role_2016}. This cosmic-ray induced chemistry was recently studied theoretically by   \citet{shingledecker_general_2018}, \citet{shingledecker_simulating_2019}, and \citet{shingledecker_cosmic-ray-driven_2018} who developed and expanded upon the idea that cosmic rays can cause ionization and excitation within the ice surface and mantle, thereby generating short-lived "suprathermal species" - a term for electronically excited molecules.  These molecules can  react with nearest neighbors either on the ice surface, or within the bulk ice, overcoming large reaction barriers, a constraint that makes thermal reactions with barriers typically slow at temperatures under 20 K. These ``suprathermal'' species either react or are quenched, and have been shown to enhance the abundances of certain species in low temperature sources, including methyl formate \citep{shingledecker_cosmic-ray-driven_2018}. Under some circumstances, photodissociation of molecules on ices to form reactive radicals can be a competitive mechanism \citep{Jin_Garrod_2020}.

In this paper, we  use the approach undertaken previously by \citet{shingledecker_cosmic-ray-driven_2018} to study the abundances of the three isomers and dimethyl ether with and without radiolysis in cold sources.  We compare our results with observed abundances for methyl formate and dimethyl ether in TMC-1, L1544,  L1689, and B1-b, and predict abundances for those species not yet observed. We also study why radiolysis can enhance the abundance of some COMs in low temperature sources such as the three isomers studied and  have smaller to nil effects on others, such as dimethyl ether.  In Section~\ref{sec:model}, we discuss our model, while in Section~\ref{sec:results} our results are shown.  The astrochemical implications are explained in Section~\ref{sec:analysis}, while Section~\ref{sec:conclusions} contains our conclusions.


\section{Model}
\label{sec:model}


\subsection{Nautilus}
\label{sec:nautilus}

The chemical simulations reported in this paper use the three-phase rate equation-based gas-grain code \texttt{Nautilus-1.1} \citep{ruaud_gas_2016}. A three-phase model is one which distinguishes species upon the surface of the grain ice (two monolayers in this case) from the remaining bulk. This distinction is relevant to diffusive chemistry, reactive desorption, and other desorption mechanisms that drive COMs from the surface into the surrounding gas, where they are more easily observed. In addition to the base \texttt{Nautilus-1.1} program, other features have been added to allow for the inclusion of radiolysis and suprathermal chemistry, as described in \citet{shingledecker_general_2018}. Radiolysis chemistry will be further described in Section~\ref{sec:radiolysis}. The chosen initial  abundances are displayed in Table~\ref{tab:abundances}, with relevant constant and homogeneous physical conditions in Table~\ref{tab:physicalcond}. Relevant to later sections is the cosmic ray ionization rate ($\zeta$), $1.3 \times 10^{-17} s^{-1}$. Table~\ref{tab:surfaceparameters}  lists the binding energies on water ice for C$_{2}$H$_{4}$O$_{2}$ isomers and dimethyl ether. The design of the \texttt{Nautilus} program allows the user to enable or disable a variety of mechanisms related to gas- and grain-processes \citep{ruaud_gas_2016}. These mechanisms and their status in this study are itemized in Table S4 of the supplementary material, entitled Nautilus Switches.


In addition to the switches listed in supplementary Table S4, others have been added to update \texttt{Nautilus} with more options.  A competitive tunneling mechanism for diffusive species, labelled "Modified\_tunneling" is described in \citet{smith_chemistry_2008}.  This choice is set to the faster option. The option "Only\_light\_tunnel" prevents all species but H and H$_2$ from tunneling under activation energy barriers of surface reactions, and is set to "on" \citep{shingledecker_simulating_2019}. There are switches regarding nonthermal desorption mechanisms for surface species. These include reactive desorption at 1 per cent probability \citep{garrod_non-thermal_2007}, photodesorption \citep{bertin_indirect_2013}, and cosmic ray induced photodesorption \citep{hasegawa_new_1993}. These three switches are set to "on", with the photodesorption yield set at $1 \times 10^{-4}$.

\begin{table}
    \centering
    \caption{Initial abundances of elements with respect to total hydrogen nuclei}
    \begin{tabular}{lcr}
        \hline
         Species & Abundance \\
        \hline
         H$_{2}^{\textup{a}}$ & $0.499$ \\
         He$^{\textup{a}}$ & $9.000 \times 10^{-2}$ \\
         N$^{\textup{a}}$ & $6.200 \times 10^{-5} $ \\
         C$^{\textup{b}}$ & $1.700 \times 10^{-4} $ \\
         O$^{\textup{c}}$ & $2.429 \times 10^{-4} $ \\
         S$^{\textup{d}}$ & $8.000 \times 10^{-8} $ \\
         Na$^{\textup{d}}$ & $2.000 \times 10^{-9} $ \\
         Mg$^{\textup{d}}$ & $7.000 \times 10^{-9} $ \\
         Si$^{\textup{d}}$ & $8.000 \times 10^{-9} $ \\
         P$^{\textup{d}}$ & $2.000 \times 10^{-10} $ \\
         Cl$^{\textup{d}}$ & $1.000 \times 10^{-9} $ \\
         Fe$^{\textup{d}}$ & $3.000 \times 10^{-9} $ \\
         F$^{\textup{e}}$ & $6.680 \times 10^{-9} $ \\
        \hline
        $^{\textup{a}}$ \citep{wakelam_polycyclic_2008} \\
        $^{\textup{b}}$ \citep{jenkins_unified_2009} \\
        $^{\textup{c}}$ \citep{mcguire_detection_2018} \\
        $^{\textup{d}}$ \citep{graedel_graphical_1982} \\
        $^{\textup{e}}$ \citep{neufeld_chemistry_2005} \\
    \end{tabular}
    \label{tab:abundances}
\end{table}

\begin{table}
    \centering
    \caption{Physical conditions for TMC-1 utilized}
    \begin{tabular}{lccr}
        \hline
         Parameter & TMC-1 \\
        \hline
         $\textit{n}_{\textup{H}}$ (cm$^{-3}$) & $10^{4}$ \\
         $\textit{n}_{\textup{dust}}$ (cm$^{-3}$) & $1.8 \times 10^{-8}$ \\
         $\textit{T}_{\textup{gas}}$ (K) & $10$ \\
         $\textit{T}_{\textup{dust}}$ (K) & $10$ \\
         $\textit{N}_\textup{{site}}$ (cm$^{-2}$) & $1.5 \times 10^{15}$ \\
         $\zeta$ (s$^{-1}$) & $1.3 \times 10^{-17}$ \\
        \hline
    \end{tabular}
    \label{tab:physicalcond}
\end{table}

\subsection{Network}
\label{sec:network}

The \texttt{Nautilus} reaction network used in this study has been expanded to include more complex chemical species and intermediates that lead up to C$_{2}$H$_{4}$O$_{2}$ isomers, as well as dimethyl ether. The base network of gaseous reactions is taken from the KIDA network \citep{wakelam_kinetic_2012}. The granular reaction network is from the \texttt{Nautilus} package, with additional thermal grain-surface reaction pathways leading to C$_{2}$H$_{4}$O$_{2}$ isomer precursors  from \citet{garrod_formation_2006}. Used initially in hot core models, these reactions have been included here to provide likely thermal pathways to COMs in competition with radiolysis. Other notable additions include new gas-phase reactions for acetic acid and glycolaldehyde from \citet{skouteris_genealogical_2018}. The full list of additional reactions is contained in the supplementary material, Tables S1 S2, and S3 (available online), with important reactions discussed in Sections~\ref{sec:methylformate}, \ref{sec:glycolaldehyde}, and \ref{sec:aceticacid}. Species with the first letter ``J''  are those on the surface of the ice mantle, while species with a ``K'' lie  within the bulk of the ice, which, after the simulation has completed running ($10^7$ yrs), contains close to 100 monolayers of ice.  
Additional binding energies $E_{\rm D}$ for the four principal molecules studied were taken from \citet{garrod_three-phase_2013} and are listed in  Table~\ref{tab:surfaceparameters}.  The diffusion barriers ($E_{\rm b}$) for all granular species are $0.4 \times E_{\rm D}$ for ice surface species, and $0.8 \times E_{\rm D}$ for bulk species.

\begin{table}
    \centering
    \caption{Binding Energies}
    \begin{tabular}{lc}
        \hline
         Species & E$_{b}$ (K)  \\
        \hline
         JHCOOCH$_{3}$ & 5200$^{a}$   \\
         JHCOCH$_{2}$OH & 6680$^{a}$  \\
         JCH$_{3}$COOH & 6300$^{a}$  \\
         JCH$_{3}$OCH$_{3}$ & 3675$^{a}$ \\
        \hline
        a \citep{garrod_three-phase_2013} \\
    \end{tabular}
    \label{tab:surfaceparameters}
\end{table}

\subsection{Radiolysis}
\label{sec:radiolysis}

The method used for radiolysis was first developed by \citet{shingledecker_general_2018}.  It describes four different results that follow an incoming cosmic ray or secondary electron interacting initially with a target species A, for which we use a  ``$\rightsquigarrow$''. The subsequent results  can be described by the following equations, which represent different paths to ions and suprathermal neutrals, the latter denoted with an asterisk:

\begin{equation}
    \mathrm{A \rightsquigarrow A^{+} + e^{-}}
    \label{eq:rad1}
\end{equation}

\begin{equation}
    \mathrm{A \rightsquigarrow A^{+} + e^{-} \rightarrow A ^{*} \rightarrow B^{*} + C^{*}}
    \label{eq:rad2}
\end{equation}

\begin{equation}
    \mathrm{A \rightsquigarrow A^{*} \rightarrow B + C}
    \label{eq:rad3}
\end{equation}

\begin{equation}
    \mathrm{A \rightsquigarrow A^{*}}
    \label{eq:rad4}
\end{equation}

\noindent    The first process leads to the ionisation of A, while the second process describes the formation of suprathermal fragments B and C with an intermediate suprathermal A formed via dissociative recombination of the A ion.  The third process describes the direct formation of an intermediate suprathermal species followed by its fragmentation into non-excited products, while the last process describes the formation of suprathermal A, which does not subsequently fragment.

In order to implement both generation of suprathermal species, and fast suprathermal reactions into the rate-equation based model used, we can derive a pseudo first-order rate coefficient by multiplying a cross section ($\sigma_e$), in cm$^{2}$, with a cosmic ray flux ($\phi$) ,in cm$^{-2}$ s$^{-1}$. In order to obtain a cross section, we start with a stopping power cross section $S(E)$, which is defined through the loss of kinetic energy of a projectile with distance and is in units of eV cm$^{2}$.  It can be calculated using the Bethe equation \citep{bethe_bremsformel_1932,johnson_energetic_1990}:

\begin{equation}
    S_{e}(E) = \frac{4 \pi Z_{x} Z_{y} e^{4}}{m_{e} \nu^{2}} \left[ \ln{\left( \frac{2m_{e} \nu^{2}}{E_{ion}} \right)} - 1 - \frac{C}{Z_{y}} \right]
    \label{eq:bethe}
\end{equation}

\noindent where $Z_{x}$ and $Z_{y}$ are the atomic numbers of the target species, and the incoming particle, respectively, $e$ is the charge of an electron, $m_{e}$ is electron mass, and $v$ is the velocity of the incoming particle. The equation refers to  the electronic stopping cross section, which is the energy lost per unit area for inelastic collisions with electrons, or cosmic ray protons. At cosmic ray energies, nuclear stopping cross sections are multiple orders of magnitude lower compared with electronic stopping cross sections; therefore, inelastic electron collisions dominate. These electronic stopping cross sections can be easily calculated using the online program \texttt{PSTAR}, which uses the Bethe equation for particles at high energies, and fits to experimental data for lower energy particles \citep{berger_estar_1999}.

From an electronic stopping cross section, we can then approximate an electronic cross section by dividing $S$ by the average energy loss per ionization for an inelastic collision ($\Bar{W}$). The average energy lost is a sum of all the processes that can result from the inelastic collisions of cosmic rays or secondary electrons with the target. These processes include the energy lost to excitation ($\Bar{W_{exc}}$), the energy loss from the generation of secondary electrons ($\Bar{W_{s}}$), and the ionization energy ($E_{ion}$), which is the average energy loss of ionization ($\Bar{W_{ion}}$) minus the energy of the ionized electron ($\epsilon$). Thus, the total average energy lost per ionization due to inelastic collisions can be defined as follows: 

\begin{equation}
    \Bar{W} = E_{ion} + \xi \Bar{W_{exc}} + \Bar{W_{s}}
    \label{eq:wbar}
\end{equation}

\noindent where $\xi$ is the average number of excitations per ionization. The value for $\Bar{W}$ is approximately 27 eV for water, but can range from 60 eV at low energy collisions to 20 eV for high energy collisions above 100 eV \citep{dalgarno_energy_1958}, and will be similar for most other molecules present in interstellar ices. For this reason, along with lack of measured values for other molecules, the average energy loss is estimated to be that of water for all reactions.

The other factor needed to calculate the quasi-first-order rate coefficient is the cosmic ray flux. We use the \citet{spitzer_heating_1968} cosmic ray distribution, integrated over their cosmic ray energy range, which \citet{shingledecker_general_2018} calculated to result in a flux ($\phi_{st}$) of 8.6 particles cm$^{-2}$ s$^{-1}$. \citet{shingledecker_general_2018} also introduced a scaling factor, $\zeta$, (with units of inverse seconds), based on the cosmic ray ionization rate of hydrogen, in order to easily adapt models to areas of higher or lower cosmic ray bombardment rates. Our overall flux is then:

\begin{equation}
    \phi = \phi_{st} \frac{\zeta}{10^{-17}}
    \label{eq:flux}
\end{equation}

\noindent The product of  $\phi$ and $\sigma_{e}$ yields the needed quasi-first-order rate coefficient:

\begin{equation}
    k = \sigma_{e} \phi = \frac{S_{e}}{\Bar{W}} \phi_{st} \frac{\zeta}{10^{-17}}
    \label{eq:rateconst}
\end{equation}

This resulting rate coefficient must be further separated into specific rate coefficients for Reaction ~\ref{eq:rad1}, ~\ref{eq:rad2}, ~\ref{eq:rad3}, and ~\ref{eq:rad4}. However, we assume the probability of Reaction~\ref{eq:rad1} to be 0, as the resulting electron will either immediately recombine, or be high enough in energy to be a reactive secondary electron. Therefore, the network does not take into account reactions similar to Reaction~\ref{eq:rad1}. The resulting reactions that are included in the chemical network, Reactions~\ref{eq:rad2}, ~~\ref{eq:rad3}, and ~\ref{eq:rad4} are referred to in \citet{shingledecker_cosmic-ray-driven_2018} as Type I, II, and III reactions, respectively.

In order to account for the differences in rates for the different ionizations or excitations, it is helpful to use the G value, which is simply a measure of the extent of a process occurring with an input of 100 eV from the cosmic ray or secondary electron.  G values can be used instead of the specific average energy loss, as shown in Equation~\ref{eq:gtow}:

\begin{equation}
    G = \frac{100 eV}{\Bar{W}}.
    \label{eq:gtow}
\end{equation}

\noindent We can estimate G values from ionization energies and average energy loss to excitation using the Shingledecker-Herbst method. \citep{shingledecker_general_2018}.
This substitution results in the final form of our equation used to obtain pseudo-first order rate coefficients for the formation of suprathermal species.

\begin{equation}
    k_{\rm n}= S_{e} \left( \frac{G_{\rm n}}{100 eV} \right) \phi_{ST} \left( \frac{\zeta}{10^{-17}} \right).
    \label{eq:kfinal}
\end{equation}

\noindent  The W values for excitation and ionization plus the ionization energies $E_{\rm ion}$ used to obtain G values are shown in Table~\ref{tab:gcalc}. The G values for reaction types I, II, and III for the COMs studied in the paper are shown in Table~\ref{tab:gvalues}. 

The suprathermal species rate coefficients are then added to the network. The result of the inclusion of suprathermal species is multifaceted. By  the generation of  suprathermal species, the model  has access to nearly instantaneous reactions that easily surmount activation energy barriers. In addition to this effect, the speed of these reactions is also enhanced because they occur between a suprathermal species and its adjacent neighbors, negating often slow diffusion processes.  Another outcome of a suprathermal species is quenching. In this process, the energy of the excited suprathermal species is vibrationally lost to the surface of the grain. The lifetime of the suprathermal species is not long enough for diffusive reactions to occur, so  rapid reactions with an adjacent neighbor or quenching can remove the excess energy of the suprathermal species. These two processes happen independently of each other, so that the rate of destruction for suprathermal species is the sum of the rates of fast reactions and quenching.  Equations for the rates of both rapid processes are given in \citet{shingledecker_cosmic-ray-driven_2018}. Despite this fast quenching process, the existence of suprathermal reactions  often allows new routes to  COMs to be enabled, even at low temperatures.

\begin{table}
    \centering
    \caption{Parameters used to estimate G values. }
    \begin{tabular}{lcccr}
        \hline
         Species & $E_{\rm ion}^{a}$ (eV) & $W_{\rm exc}^{b}$ (eV) & W$_{\rm s}$ (eV)  \\
        \hline
         HCOOCH$_{3}$ & 10.84 & 5.390 & 3.636 \\
         HCOCH$_{2}$OH & 10.86 & 4.132 & 3.639 \\
         CH$_{3}$COOH & 10.65 & 6.199 & 3.614 \\
         CH$_{3}$CHO & 10.23 & 4.025 & 3.563 \\
         HCO & 8.12 & 2.017 & 3.256 \\
        \hline
        $^{a}$ \citep{lias_ion_2018} \\
        $^{b}$ \citep{keller-rudek_mpi-mainz_2013}
        
    \end{tabular}
    \label{tab:gcalc}
\end{table}

\begin{table}
    \centering
    \caption{Estimated G values (extent per 100 eV)}
    \begin{tabular}{lccc}
        \hline
         Species & G$_{\rm I}$ & G$_{\rm II}$ & G$_{\rm III}$ \\
        \hline
         HCOOCH$_{3}$ & 3.704 & 4.305 & 4.305 \\
         HCOCH$_{2}$OH & 3.704 & 5.603 & 5.603 \\
         CH$_{3}$COOH & 3.704 & 3.804 & 3.804 \\
         CH$_{3}$CHO & 3.704 & 6.076 & 6.076 \\
         HCO & 3.704 & 14.345 & 14.345 \\
        \hline
    \end{tabular}
    \label{tab:gvalues}
\end{table}

\section{Results and Analysis}
\label{sec:results}


In our calculations, we have utilized two low temperature models, numbered 1 and 2. Model 1 does not contain radiolysis chemistry, and results with this model will be represented by solid lines in figures.   Model 2 includes radiolysis chemistry, resulting in the generation of suprathermal species. Results with this model are shown in figures with dashed lines. Both models utilize the same set of choices regarding \texttt{Nautilus} switches, except for the switch that enables radiolysis. The models likewise share the same initial abundances and physical parameters.  

 New sets of reactions have been included in both of our models.  New radiolysis reactions have been included to Model 2, which can increase the production of suprathermal species. Notable in Model 2 are not just the new suprathermal reactions that quickly produce the major molecules in our study, but also additional such reactions that destroy these molecules, which must be closely examined. 
 Without these reactions, it is possible to overproduce the COMs we are studying in a non-physical manner. Model 1 includes the new thermal production and destruction reactions that are relevant to the species examined here that are also in Model 2.

 In the modifications we have made to the networks, notable reactions include all production and destruction reactions involving acetic acid and glycolaldehyde on grains, as there are no reactions , thermal or suprathermal,  that are included in the standard reaction network used.  For example, in Model 2, we have included suprathermal hydrogen abstraction reactions that reduce  double bonded oxygen atoms to hydroxy (-OH) groups.  These reactions destroy acetic acid and glycolaldehyde. and generate suprathermal species. Also included are other new rapid reactions involving suprathermal species. However, the destruction reactions of more complex species, not only for glycolaldehyde and acetic acid, have not been  examined sufficiently, as most efforts to examine them have been focused on production routes. A detailed consideration of such destruction pathways have been found to be useful in previous investigations of families of isomers \citep{shingledecker_case_2019,shingledecker_isomers_2020}, however such a study is beyond the scope of this work. Additionally, although methyl formate and dimethyl ether are both included in the base KIDA gas-phase network, glycolaldehyde and acetic acid are not,  meaning there are very few thermal gas phase routes leading to their formation, and even fewer destruction reactions involving them, both in the gas phase and solid phase of the ice on the dust grains. This paucity of reactions must be eliminated.

\subsection{Methyl Formate}
\label{sec:methylformate}

In warmer environments, the main method of producing methyl formate is likely through a series of  diffusive grain surface reactions \citep{laas_contributions_2011}. The final reaction in the chain is:

\begin{equation}
    \mathrm{JHCO + JCH_{3}O \rightarrow HCOOCH_{3}},
    \label{eq:mf1}
\end{equation}

\noindent after which the methyl formate product is subsequently desorbed from the grain surface once the temperature reaches $\approx$~100~K. At lower temperatures, desorption occurs mainly through the exothermicity of the reaction 1 per cent of the time.  A series of gas phase reactions also leads to gaseous methyl formate with the final reaction involving atomic oxygen and a radical \citep{balucani_formation_2015}:

\begin{equation}
    \mathrm{CH_{3}OCH_{2} + O \rightarrow HCOOCH_{3} + H},
    \label{eq:mfgas}
\end{equation}

\noindent although the role of this reaction has mainly been studied at low temperatures. 

Although the abundance of methyl formate in warm regions is reproduced satisfactorily by the approach of thermal diffusion reactions followed by thermal desorption, the situation is quite different at temperatures below 20 K. In addition to the difficulty of producing the molecule on grains, desorption can only occur by non-thermal means, such as reactive desorption. At these temperatures,  methyl formate has been detected toward various sources with cold and pre-stellar cores, with approximate fractional abundances with respect to hydrogen of $7.4 \times 10^{-10})$ toward L1689B, $1.5 \times 10^{-10}$ toward TMC-1,  and $2.0 \times 10^{-11}$ toward B1-b \citep{bacmann_detection_2012,cernicharo_discovery_2012,soma_complex_2018} . 

\begin{figure*}
    \centering
	\includegraphics[width=\textwidth]{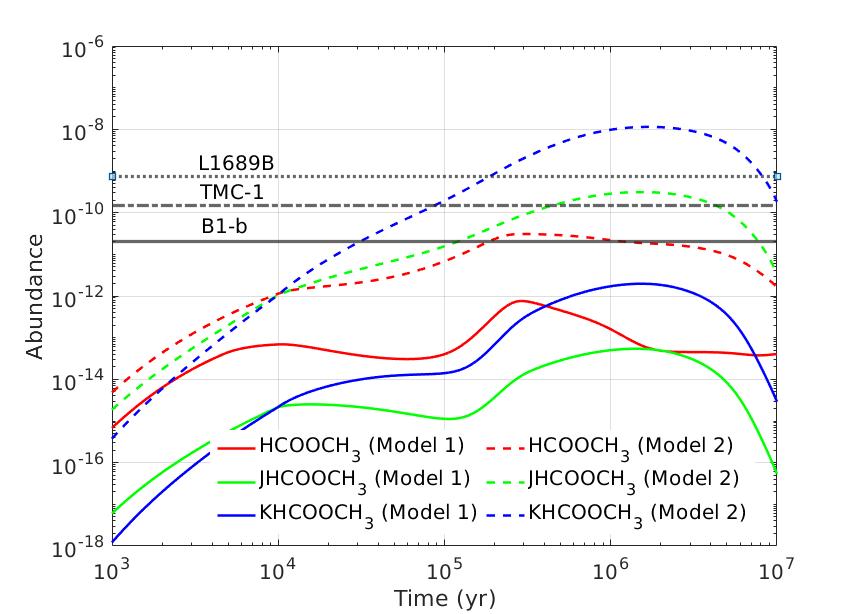}
    \caption{ \large Abundance of methyl formate. Gas phase is in red, ice surface in green, and ice bulk in blue. Both models have the same physical and starting conditions, and are run at 10 K.  The solid lines refer to the model without radiolysis (Model 1), while the dashed lines refer to the model with radiolysis turned on (Model 2). Horizontal black lines refer to observed abundances in L1689B, TMC-1, and B1-b.}
    \label{fig:mfabundance}
\end{figure*}

Let us first consider how the results of our non-radiolysis model (Model 1) compare with these observations.  As shown in Figure~\ref{fig:mfabundance}, the predicted peak fractional abundance of gaseous methyl formate is only ${7.5 \times 10^{-13}}$ at $\approx 3 \times 10^{5}$ yr whereas the  observed value in B1-b is still a factor of 30 higher, that in TMC-1 a factor of 200 higher, and that in L1689  three orders of magnitude higher. Much better results are found in our model with radiolysis, which achieves a value of $3 \times 10^{-11}$, at a time of  $2.75 \times 10^{5}$ yr, as can be seen in the figure, and will be discussed below. Nevertheless, even the result of Model 2 is more than an order of magnitude below the abundance in L1689B, although quite adequate for the other two sources.

Why does Model 1 fail to reproduce observed methyl formate abundances at 10 K. At this temperature,  the main limitation to Reaction~(\ref{eq:mf1}) is not the presence of an activation energy barrier, but  diffusion barriers for both species \citep{garrod_formation_2006}. The radicals HCO and CH$_{3}$O are too massive to reliably tunnel under the diffusion barriers of 1600 K and 5080 K, respectively, and cannot overcome the barrier by hopping because of the low temperature \citep{garrod_formation_2006}. Therefore, in order to reproduce the observed abundances of methyl formate in parts of the ISM at 10 K, other methods are needed. Radiolysis is the method pursued here.

\citet{shingledecker_cosmic-ray-driven_2018} used the \texttt{Nautilus-1.1} model to show how radiolysis increases the abundance of gaseous methyl formate, among other species, in TMC-1 and similar sources \citep{ruaud_gas_2016}. Fig.~\ref{fig:mfabundance} corroborates their results.  In Model 2 and in their work, a peak abundance slightly greater than 10$^{-11}$ is achieved at a time of $ \approx 3 \times 10^{5}$ yr, while Model 2 has a larger abundance than their value at times earlier than $\sim 10^{5}$ yr.  While all reactions examined in this section are included in \citet{shingledecker_cosmic-ray-driven_2018}, we have  included additional thermal production and destruction mechanisms as well as new radiolysis destruction mechanisms for methyl formate. These newly included reactions are identified in the supplementary material (available online), and are not the dominant reactions that produce and destroy methyl formate  \citep{shingledecker_cosmic-ray-driven_2018}. However, the somewhat larger abundance of methyl formate in Model 2 at non-peak times  is caused by the newly included reactions.

The  dominant surface reaction scheme to form methyl formate was initially proposed by \citet{bennett_laboratory_2005}, and starts with the electron-irradiation of methanol and carbon monoxide ices, which lead to the production of  both methyl formate and glycolaldehyde, depending upon whether the molecule reacting with HCO is methoxy (CH$_{3}$O), the precursor of methyl formate, or hydroxymethyl (CH$_{2}$OH), the precursor of glycolaldehyde. These reactions are included in \citet{shingledecker_cosmic-ray-driven_2018}. In our Model 2, either HCO or CH$_{3}$O can be suprathermal in order to produce methyl formate efficiently as shown in the following reactions:

\begin{equation}
    \mathrm{JHCO^{*} + JCH_{3}O \rightarrow HCOOCH_{3}},
    \label{eq:mf2}
\end{equation}

\begin{equation}
    \mathrm{JHCO + JCH_{3}O^{*} \rightarrow HCOOCH_{3}}.
    \label{eq:mf3}
\end{equation}

\noindent Indeed, reactions~\ref{eq:mf2} and \ref{eq:mf3} are the two main reactions that produce methyl formate in our radiolysis model - Model 2. Reaction~\ref{eq:mf3} is slightly more important, due to the higher abundance of CH$_{3}$O present in the ice, which contributes to a higher number of suprathermal CH$_{3}$O molecules being generated. There are multiple gas-phase formation routes, but we find these reactions to be slower than the suprathermal reactive desorption reactions above, which come directly from radiolysis. 

For methyl formate, unlike the case for the other species in this study, there are already adequate  destruction mechanisms both in the gas phase and for the ice surface and bulk. The main destruction pathways in both models for gaseous methyl formate are 

\begin{equation}
    \mathrm{HCOOCH_{3} + C \rightarrow CO + CO + H + CH_{3}}
    \label{eq:mfdest1}
\end{equation}

\begin{equation}
    \mathrm{HCOOCH_{3} + H_{3}^{+} \rightarrow H_{2} + H_{5}C_{2}O_{2}^{+}}
    \label{eq:mfdest2}
\end{equation}

 Reaction~\ref{eq:mfdest1} is dominant until approximately $10^6$ yr while Reaction~\ref{eq:mfdest2} is dominant after this time, while still being significant earlier. This difference arises from the time dependence of the abundance of gas phase C and H$_3^+$; C is mostly present at early times until it gets trapped in other carbon-bearing species at later stages,
 even in purely gas-phase models. In contrast, the abundance of H$_{3}^{+}$ is consistently lower than C before $10^{6}$ yr with an abundance of  $ \approx 4 \times 10^{-10}$.  After $10^{6}$ yr, the abundance of H$_{3}^{+}$ nearly matches that  of gaseous carbon  in both models, with abundances of $ \approx 1.5 \times 10^{-8}$.

The calculated abundances of  methyl formate in the ice phases as functions of time are also shown in Figure~\ref{fig:mfabundance}. At the current stage of infrared astronomy, however, it is not possible to compare the calculated ice abundances with observational results for a number of reasons, such as the difficulty of observing weak and broad features in space. With the eventual launch of the James Webb Space Telescope, more information on ice composition may become available.

\subsection{Acetic Acid}
\label{sec:aceticacid}

Acetic acid has been detected toward SgrB2, but has yet to be detected toward colder regions of the ISM \citep{mehringer_detection_1997,remijan_acetic_2002,xue_alma_2019,xue_alma_2019-1}. Prior to this work, as far as we can determine, the abundance of acetic acid toward cold areas of the ISM has not been included in granular chemistry in previous modelling attempts of cold dark clouds.
 
Numerous gas phase methods have been proposed for assorted COMs, most involving ion-neutral reactions followed by dissociative electron recombination \citep{huntress_synthesis_1979,ehrenfreund_organic_2000,blagojevic_gas-phase_2003}.  As far as we know, very few of these suggestions involve the formation of gaseous acetic acid.  However, recently a quantum chemical study was performed on the gas phase synthesis of acetic acid and glycolaldehyde  originating from ethanol \citep{skouteris_genealogical_2018}. These reactions have been included in our treatment; however,  results from Model 2 suggest that 1 per cent reactive desorption from grain ices produces the majority of acetic acid through the following grain surface reactions at 10 K:

\begin{equation}
    \mathrm{JHOCO + JCH_{3} \rightarrow CH_{3}COOH}
    \label{eq:aa1}
\end{equation}

\begin{equation}
    \mathrm{JOH^{*} + JCH_{3}CO \rightarrow CH_{3}COOH}.
    \label{eq:aa2}
\end{equation}

\noindent Reaction~\ref{eq:aa1} is a standard thermal reaction from \citet{garrod_complex_2008}, while reaction~\ref{eq:aa2} is a suprathermal version of a standard thermal reaction from the same source. One of the gas phase reactions from \citet{skouteris_genealogical_2018} contributes in the production of acetic acid at 10 K via the reaction

\begin{equation}
    \mathrm{CH_{3}CHOH + O \rightarrow CH_{3}COOH + H.}
    \label{eq:aagas}
\end{equation}
\noindent This gas-phase reaction contributes efficiently for a short period of time between $1.5 \times 10^{5}$ yr and $3 \times 10^{5}$ yr, but is less efficient than Reactions~\ref{eq:aa1} and ~\ref{eq:aa2} at all other times in the model.

\begin{figure*}
    \centering
	\includegraphics[width=\textwidth]{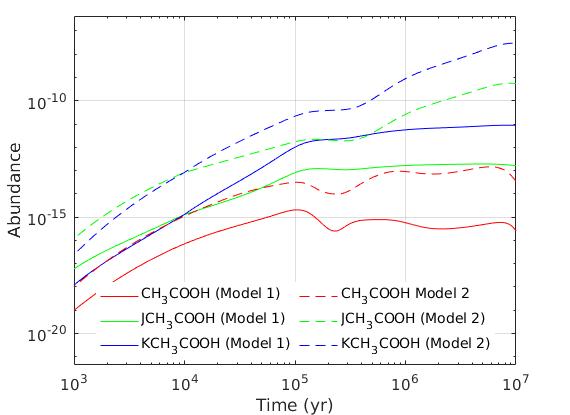}
    \caption{ \large Abundance of acetic acid. Gas phase is in red, ice surface in green, and ice bulk in blue. Both models have the same physical and starting conditions, and are run at 10 K. The solid lines show the model without radiolysis, while the dashed lines show the model with radiolysis turned on.}
    \label{fig:aaabundance}
\end{figure*}

There may need to be further examination of Reaction~\ref{eq:aa1}, because the reaction of HOCO with CH$_{3}$ might require one of the reactants to be a suprathermal species to overcome an activation energy barrier \citep{bennett_formation_2007}. Surprisingly, in Model 2,however, the thermal Reaction~\ref{eq:aa1} is more efficient than the same reaction involving suprathermal species.  This effect could be due to the lack of HOCO* and CH3* at relevant times. The reactions lead to a maximum gas-phase abundance in Model 2 of $1.375 \times 10^{-13}$  as shown in Fig.~\ref{fig:aaabundance}, compared with the peak abundance from Model 1 of $1.86 \times 10^{-16}$. The higher calculated abundance is not sufficient for current telescopes to detect acetic acid towards TMC-1.

In comparison with methyl formate and dimethyl ether, there are a limited number of gas-phase acetic acid destruction reactions,  all involving ions. Reactions taken from \citet{skouteris_genealogical_2018} suggest that gas phase destruction reactions of acetic acid  are mostly through ion-neutral reactions such as Reactions~\ref{eq:aagasdest1}, ~\ref{eq:aagasdest2}, and ~\ref{eq:aagasdest3}:

\begin{equation}
    \mathrm{ H_{3}O^{+} + CH_{3}COOH  \rightarrow CH_{3}COOH_{2}^{+} + H_{2}O },
    \label{eq:aagasdest1}
\end{equation}

\begin{equation}
    \mathrm{ H_{3}^{+} + CH_{3}COOH  \rightarrow CH_{3}COOH_{2}^{+} + H_{2} },
    \label{eq:aagasdest2}
\end{equation}

\begin{equation}
    \mathrm{ H_{3}^{+} + CH_{3}COOH \rightarrow CH_{3}CO^{+} + H_{2} + H_{2}O },
    \label{eq:aagasdest3}
\end{equation}

These reactions are distinct from the neutral-neutral reactions that destroy methyl formate more efficiently, such as the destruction reaction involving atomic C (Reaction~\ref{eq:mfdest1}), which acetic acid does not have in our network. Both Model 1 and Model 2 show production and destruction rates for methyl formate at least four orders of magnitude faster than the production and destruction routes for acetic acid.

\subsection{Glycolaldehyde}
\label{sec:glycolaldehyde}

\begin{figure*}
    \centering
	\includegraphics[width=\textwidth]{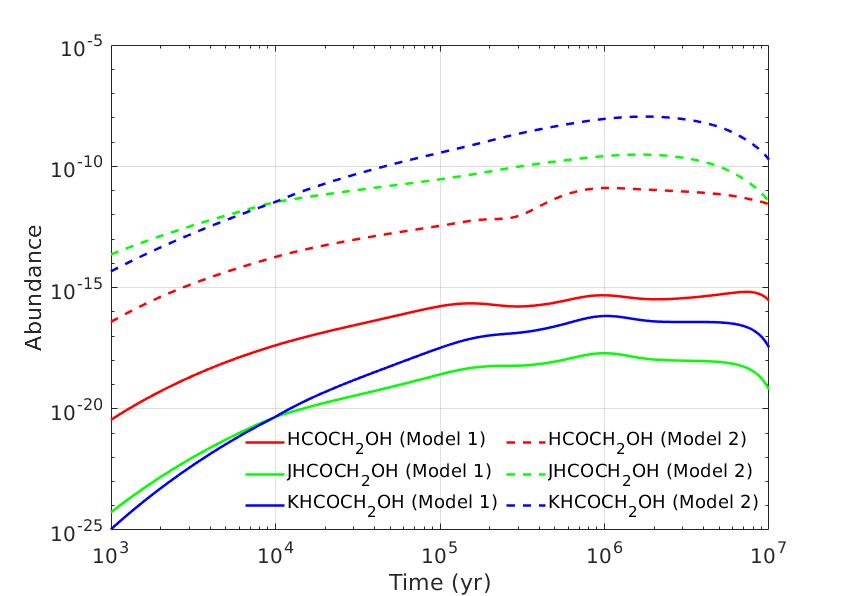}
    \caption{ \large Abundance of glycolaldehyde. Gas phase is in red, ice surface in green, and ice bulk in blue. Both models have the same physical and starting conditions, and are run at 10 K. The solid line shows the model with radiolysis off, while the dashed line is the model with radiolysis turned on.}
    \label{fig:gabundance}
\end{figure*}

Like acetic acid, glycolaldehyde has no reactions in the KIDA network.  We have included thermal grain surface and ice mantle reactions from \citet{garrod_formation_2006},  \citet{hudson_ir_2005}, and \citet{garrod_complex_2008}, and gas-phase reactions from \citet{skouteris_genealogical_2018}. Previous research into the synthesis of glycolaldehyde suggests that most routes previously studied  are not efficient enough to produce significant amounts in the ISM  \citep{woods_formation_2012,woods_glycolaldehyde_2013}. Even with the addition of reactions from  \citet{skouteris_genealogical_2018}, production of gas-phase glycolaldehyde is still lacking without radiolysis, as can be seen in Fig.~\ref{fig:gabundance}. The inclusion of radiolysis greatly enhances the abundance of glycolaldehyde with the abundance peaking at about the same abundance as methyl formate, albeit at later times in the model. The peak abundance of gaseous glycolaldehyde is about $1.3 \times 10^{-11}$ with radiolysis, increasing from the non-radiolysis abundance of about $5.0 \times 10^{-16}$ . The main reactions that produce gas-phase glycolaldehyde are


\begin{equation}
    \mathrm{JCH_{2}OH + JHCO^{*} \rightarrow HCOCH_{2}OH},
    \label{eq:g2}
\end{equation}
and
\begin{equation}
    \mathrm{JCH_{2}OH^{*} + JHCO \rightarrow HCOCH_{2}OH}.
    \label{eq:g3}
\end{equation}

\noindent  Reactions~\ref{eq:g2} and \ref{eq:g3} are grain-ice surface reactions that  lead initially to grain surface glycolaldehyde followed by reactive desorption  \citep{bennett_formation_2007-1, garrod_non-thermal_2007}. The faster reaction is Reaction~\ref{eq:g2} until about $5 \times 10^{6}$ yr, at which point Reaction \ref{eq:g3} produces the majority of gaseous glycolaldehyde. This is due to the time it takes to generate CH$_{2}$OH and CH$_{2}$OH$^{*}$.  The importance of the CH$_{2}$OH precursor is highlighted in \citet{bennett_formation_2007-1}, because the different reactions of  CH$_{2}$OH and CH$_{3}$O with HCO determine if the product will be glycolaldehyde or methyl formate.  For  CH$_{2}$OH, Models 1 and 2 lead to a peak abundance of $\sim 10^{-9}$, while CH$_{3}$O has a peak abundance of $\sim 10^{-7}$.
The resulting abundances of methyl formate and glycolaldehyde reflect the abundances of their precursors. 

Like acetic acid, glycolaldehyde lacks neutral-neutral gas phase destruction mechanisms that methyl formate and dimethyl ether possess. The destruction rates of  gas phase glycolaldehyde from \citet{skouteris_genealogical_2018} included here are so slow that our model has adsorption of glycolaldehyde on to the grains as faster. Notable is the inclusion of the hydrogen atom addition reactions \ref{eq:gadest1} and \ref{eq:gadest2} for the destruction of grain-surface glycolaldehyde : 

\begin{equation}
    \mathrm{ JHCOCH_{2}OH + JH \rightarrow JH_{2}COCH_{2}OH + JH \rightarrow J(CH_{2}OH)_{2}}
    \label{eq:gadest1}
\end{equation}

\begin{equation}
    \mathrm{ KHCOCH_{2}OH + KH \rightarrow KH_{2}COCH_{2}OH + KH \rightarrow K(CH_{2}OH)_{2}}
    \label{eq:gadest2}
\end{equation}
\noindent that result in the production of hydrogenated glycolaldehyde. Without these reactions, the amount of surface ice glycolaldehyde would be greater than the abundance of methyl formate in the ice. 
It should be mentioned at this stage that glycolaldehyde can be made efficiently on  10 K CO ice  in a thermal laboratory experiment  via a reaction sequence starting with CO and hydrogen, and thought to proceed via the dimerization of HCO followed by the formation of a C-C bond although the conditions are not the same as in the ISM  \citep{fedoseev_experimental_2015, fuchs_hydrogenation_2009}.

\subsection{Dimethyl Ether}
\label{sec:dimethyl}
 
\begin{figure*}
    \centering
	\includegraphics[width=\textwidth]{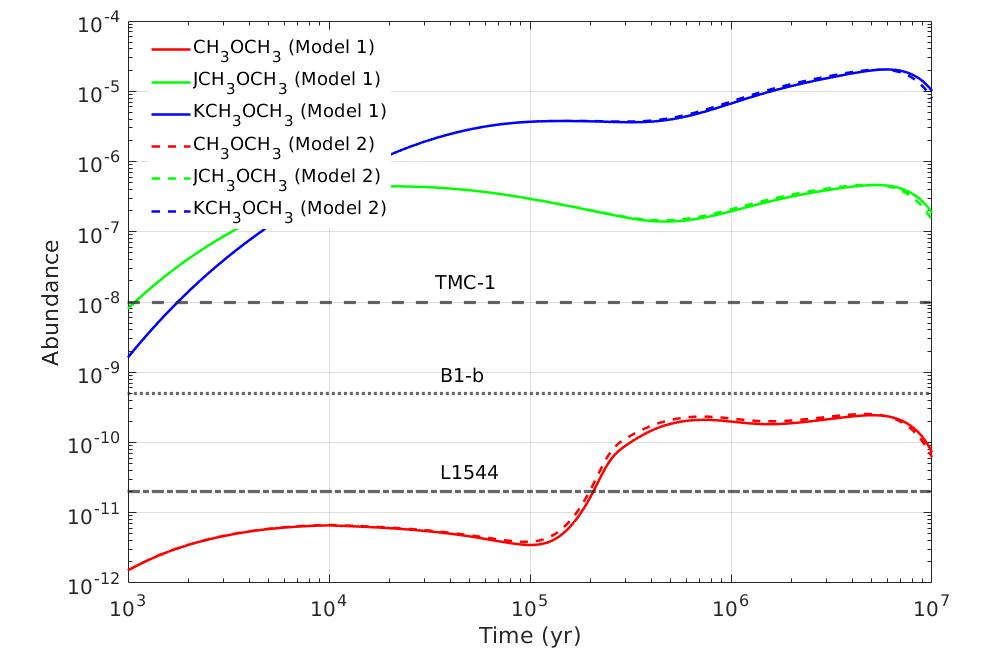}
    \caption{ \large Abundance of dimethyl ether in the gas phase (red), ice surface (green), and ice bulk (blue). Both models have the same physical and starting conditions, and are run at 10 K. The solid line shows the model without radiolysis, while the dashed line is the model with radiolysis. Horizontal black lines refer to observed abundances in TMC-1, L1544, and B1-b.}
    \label{fig:deabundance}
\end{figure*}

Dimethyl ether has been detected toward multiple cold sources, such as TMC-1, L1544, and B1-b, with reported abundances of $\approx$ $1 \times 10^{-8}$ in TMC-1, $\approx$ $2 \times 10^{-11}$ in L1544, and  $\approx$ $5 \times 10^{-10}$ in B1-b \citep{soma_complex_2018,vastel_origin_2014,jimenez-serra_spatial_2016,cernicharo_discovery_2012}. 
The formation of dimethyl ether has been closely linked to the formation of C$_2$H$_4$O$_2$ isomers, as they are thought to share similar formation routes and precursors. Dimethyl ether is unique in this paper, however, as can be seen in Figure~\ref{fig:deabundance}, which shows that radiolysis does not enhance the gas-phase abundance significantly compared with the other species examined here.   The calculated abundances in the gas phase with respect to hydrogen peak at $\approx$ $2.5 \times 10^{-10}$ with radiolysis, and approximately $2.4 \times 10^{-10}$ for the model without radiolysis. Both of these values provide a reasonable fit to observations. The small difference between M1 and M2 arises from the main production routes occurring thermally on dust grains without suprathermal reactants; e.g.,

\begin{equation}
    \mathrm{ JH + JCH_{3}OCH_{2} \rightarrow CH_{3}OCH_{3},}
    \label{eq:de1}
\end{equation}

\begin{equation}
    \mathrm{ JCH_{3} + JCH_{3}O \rightarrow CH_{3}OCH_{3}.}
    \label{eq:de2}
\end{equation}

\begin{figure}
	\includegraphics[width=\columnwidth]{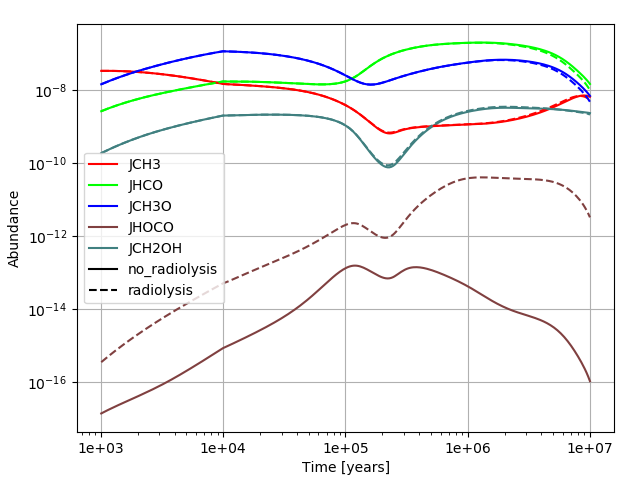}
    \caption{ Abundances of selected precursor species on the ice surface  at 10 K. The solid line refers to the model without radiolysis (Model 1), while the dashed line refers to the model with radiolysis (Model 2).}
    \label{fig:iceintermediates}
\end{figure}

\begin{figure}
    \includegraphics[width=\columnwidth]{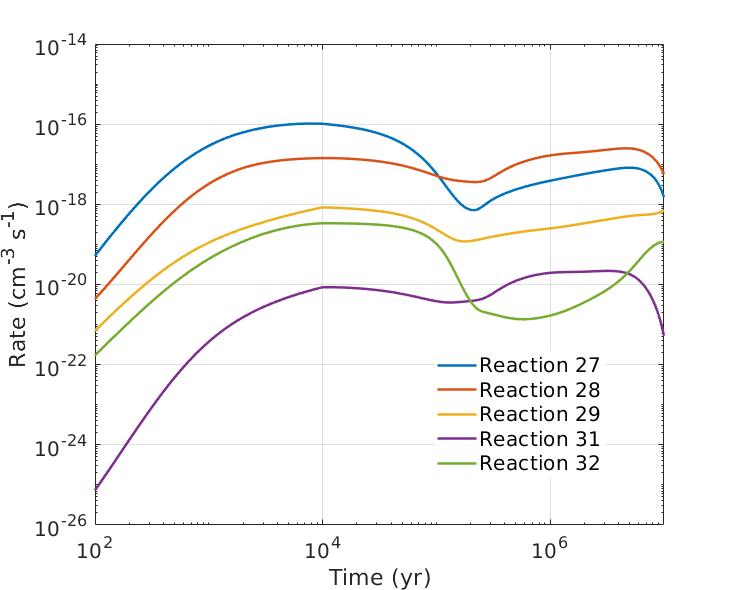}
    \caption{ Rates of relevant CH$_{3}$OCH$_{3}$ production mechanisms. Reactions 27 and 28 are in Model 1 (solid lines) and 2 (dashed lines), reactions 29, 31, and 32 are only in Model 2.}
    \label{fig:rates_prod_de}
\end{figure}

Despite the inclusion of suprathermal reactions such as the following:

\begin{equation}
    \mathrm{JH^{*} + JCH_{3}OCH_{2} \rightarrow CH_{3}OCH_{3},}
    \label{eq:de3}
\end{equation}

\begin{equation}
    \mathrm{JH + JCH_{3}OCH_{2}^{*} \rightarrow CH_{3}OCH_{3},}
    \label{eq:de4}
\end{equation}

\begin{equation}
    \mathrm{ JCH_{3}^{*} + JCH_{3}O \rightarrow CH_{3}OCH_{3},}
    \label{eq:de5}
\end{equation}

\begin{equation}
    \mathrm{ JCH_{3} + JCH_{3}O^{*} \rightarrow CH_{3}OCH_{3},}
    \label{eq:de6}
\end{equation}

\noindent none of these reactions dominates, unlike the case of the C$_2$H$_4$O$_2$ isomers, because of the efficiency of Reactions~\ref{eq:de1} and ~\ref{eq:de2}. Reaction~\ref{eq:de1} is still efficient at low temperatures because hydrogen on the grain surface is still mobile, and is more likely to move to encounter JCH$_{3}$OCH$_{2}$. In order to further explain the other reactions leading to dimethyl ether, a knowledge of the abundance of precursor molecules will be useful, as will direct comparison of rates between models and the reactions mentioned above. Fig.~\ref{fig:iceintermediates} shows the abundance of such relevant precursor molecules, while Fig.~\ref{fig:rates_prod_de} shows the rates for the reactions mentioned above. Most of these intermediates are not significantly enhanced by radiolysis chemistry, and the dashed and solid lines appear as one. 
Clearly, the abundances of both JCH$_{3}$ and JCH$_{3}$O are sufficient at relevant times that thermal methods can produce dimethyl ether at the observed levels at 10 K.  Had the switch for “is\_crid” been “on”, diffusion would have slightly increased for all species and helped to increase abundances by increasing rates of reaction. This switch is not turned on in the current model, but we are working on a more complex and accurate approach.

Although one would expect Reaction~\ref{eq:de3} to dominate due to the efficient speed of most other radiolysis reactions and the high availablilty of H$^{*}$ from water ice, Fig.~\ref{fig:rates_prod_de} shows that Reactions ~\ref{eq:de1} and ~\ref{eq:de2} dominate in models with and without radiolysis. 

Interestingly, Reaction 30 is not included in Fig.~\ref{fig:rates_prod_de} because of rates constantly below $10^{-40}$ cm$^{-3}$ s$^{-1}$. This may be caused by the slow generation of JCH$_{3}$OCH$_{2}^{*}$. In general, however, since atomic hydrogen can tunnel efficiently under diffusion barriers, allowing for very efficient surface reactions even with other heavy species at low temperatures as long as there is no activation energy, thermal routes with complex molecules and thermal atomic hydrogen can compete with suprathermal reactions at low temperatures. One similar formation mechanism is that of methanol, where models can efficiently produce methanol on dust grains even at low temperatures through the rapid movement of thermal atomic hydrogen.

\section{Astrochemical Implications}
\label{sec:analysis}

All three of the C$_2$H$_4$O$_2$ isomers studied in this paper show significant increases in calculated gas-phase abundances when radiolysis chemistry is included in our model calculations at 10 K. Of the isomers, methyl formate has already been well observed within colder environments within the ISM, and has been the subject of a number of previous simulations.  However, previous synthetic treatments have underproduced or barely produced sufficient methyl formate \citep{chang_unified_2016, balucani_formation_2015,vasyunin_unified_2012}. \citet{shingledecker_general_2018}  first showed that radiolysis greatly improves the gas-phase abundance of multiple molecules, including methyl formate. In this work, we have added new destruction methods of methyl formate through radiolysis, and thermal production and destruction mechanisms, listed in supplementary material (available online) (K. Acharyya, private communication). The molecules examined in this work show that radiolysis should enhance the abundances of more species, potentially allowing their detection in dark clouds.  One benefit of including radiolysis is to allow for fast reactions with immediate neighbors in the ice in which the suprathermal species can have enough energy to overcome most reaction barriers before being quenched. This process is in contrast with many surface radical-radical reactions at low temperatures, which generally have a low or no chemical barrier at all, but have high enough diffusion barriers to seriously hamper any reaction that is limited by thermal diffusion, both on the ice surface, and in the bulk of the ice.

Most of the reactions  examined in this paper occur between ice surface species that form products which  are  assumed to desorb given sufficient energy at 1 per cent of the products. The molecules that remain in or on the ice can lead to the production of more complex COMs, by generating  suprathermal species, or reacting with smaller species to increase complexity. For example, methyl formate that does not desorb following a reaction can undergo radiolysis, generating both suprathermal and thermal CH$_{3}$O and HCO. Alternatively, HOCO, an intermediate species that is enhanced by radiolysis ( Fig.~\ref{fig:iceintermediates}), can thermally react with either H or CH$_{3}$ to produce HCOOH or acetic acid, respectively.

The increase in abundance for the C$_{2}$H$_{4}$O$_{2}$ isomers shows that it is possible to build up abundances of complex molecules such as methyl formate, acetic acid, and glycolaldehyde at 10 K, both in the gas phase and in the ice mantle of the dust grains through radiolytic chemistry and thermal reactions. 

Despite the role of dimethyl ether as a COM, a role shared with the three isomers studied, and its detection in similar regions with methyl formate, radiolysis does not significantly impact the calculated gas-phase abundance of this molecule.  This difference arises from the efficiency of normal thermal pathways and the lack of efficient suprathermal precursors in the synthesis of dimethyl ether. This exception to the role of radiolysis in enhancing abundances of organic molecules is important to show that not all species are significantly affected positively by radiolysis chemistry. Indeed, there may be efficient thermal routes already in existence, or yet to be examined routes to even more complex molecules that do not rely solely on radiolysis, such as cyanopolyynes \citep{shingledecker_cosmic-ray-driven_2018}.

 Although gas phase reactions that generate acetic acid and glycolaldehyde have been included, they are mostly insignificant compared with the resulting rates of the suprathermal reactions. More gas phase quantum mechanical studies or experimental studies on reaction rates should be conducted on all species examined here, but acetic acid is still lacking in information compared with methyl formate, glycolaldehyde, and dimethyl ether. These studies should be done in order to potentially find new pathways, because  we could be ignoring other granular or gas-phase routes to all of these molecules.

Unlike the situation for acetic acid  and glycolaldehyde, a gas-phase synthesis for dimethyl ether and methyl formate has been constructed by \citet{balucani_formation_2015}, based at least partially on measured and calculated rate coefficients involving neutral species. An earlier treatment was reported by \citet{vasyunin_unified_2013}. One important reaction in the sequence is a radiative association between the methyl and methoxy radicals to form dimethyl ether, which has not been studied in the laboratory:
\begin{equation}
    \mathrm{ CH_{3} + CH_{3}O \rightarrow CH_{3}OCH_{3}+ h\nu.}
    \label{eq:de7}
\end{equation}
Calculations by Tennis, Loison \& Herbst (in preparation) indicate that this reaction occurs at near the collisional value at 10 K. With this large rate coefficient, the gas-phase approach of \citet{balucani_formation_2015} embedded in the OSU2009 network shows significant enhancements to the calculated abundances of dimethyl ether and methyl formate, and can account for the observation of the former in L1544.

There are certain aspects of the treatment of radiolysis that can possibly be improved. For example, the assumption that suprathermal species react with unit efficiencies should be examined more closely.
In addition, there are other aspects of cosmic ray interactions that have yet to be implemented into \texttt{Nautilus}, such as a more advanced treatment of cosmic ray grain heating, and sputtering caused by cosmic rays, both of which should help remove species present in the surface layers and the bulk of the ice \citep{kalvans_chemical_2019}. This is important because most species formed through radiolysis and other grain production routes remain adsorbed on the ice surface or trapped within the ice bulk, as shown in Figures ~\ref{fig:mfabundance}, ~\ref{fig:aaabundance},  ~\ref{fig:gabundance}, and
\ref{fig:deabundance}. 

Inclusion of low energy resonances below the minimum energy for suprathermal species has been suggested by C. Arumainayagam, private communication. Further studies are also needed in the areas of diffusion on and within the ice, because there appears to be some evidence that a diffusion-like process has been missed \citep{shingledecker_efficient_2020}. Another problem concerns the desorption of bound species from the ice surface and bulk mantle. At 10 K, not even most volatile species will be thermally desorbed rapidly; hence the importance of photodesorption, reactive desorption, and sputtering caused by cosmic rays \citep{dartois_non-thermal_2019}. There are other treatments of reactive desorption that have a more nuanced approach than the base probability of desorption based on RRK theory \citep{garrod_non-thermal_2007, minissale_dust_2016} or our assumed probability of 1 per cent; however, an examination of these treatments is outside the scope of this paper.


\section{Conclusions}
\label{sec:conclusions}

In this work, we have added to the use of the method of radiolytic chemistry via the generation of suprathermal species to determine the abundances of selected COMs in cold dark clouds, such as TMC-1. The base versions of \texttt{Nautilus-1.1} and the KIDA network have been expanded to allow for radiolysis chemistry in modeling the chemistry of the C$_{2}$H$_{4}$O$_{2}$ isomers methyl formate, glycolaldehyde, and acetic acid, as well as the important COM dimethyl ether. The results of the models discussed here show that radiolytic chemistry greatly increases the modelled abundance of these species except for dimethyl ether, so much so that it may be possible to detect glycolaldehyde in dark clouds. However, there is likely not enough gaseous acetic acid to be observed. Additionally, there are abundances of the C$_{2}$H$_{4}$O$_{2}$ isomers and dimethyl ether remaining within the ice. These remaining species could be used as building blocks for molecules of even greater complexity if activated by UV photons, as discussed by \citet{Jin_Garrod_2020} or by cosmic rays.  If the very complex molecules remain on the grain mantle, however, infrared detection will be difficult for a number of reasons, although the advent of JWST might allow some more detections.

\section*{Acknowledgements}

We acknowledge and thank the anonymous referee for his comments and careful reading. E. H. thanks the National Science Foundation (US) for support of his research programme in astrochemistry through grant AST 19-06489. C. N. S. thanks the Alexander von Humboldt Stiftung/Foundation for their generous support. This research has made use of NASA's Astrophysics Data System Bibliographic Services. We would like to thank V. Wakelam for the use of the \texttt{Nautilus-1.1} program.

\section*{Data Availability}
The data underlying this article are available in the article, in its online supplementary material, as well as cited online repositories (KIDA).




\bibliographystyle{mnras}
\bibliography{references,chris_references}

\begin{thebibliography}{}
\makeatletter
\relax
\def\mn@urlcharsother{\let\do\@makeother \do\$\do\&\do\#\do\^\do\_\do\%\do\~}
\def\mn@doi{\begingroup\mn@urlcharsother \@ifnextchar [ {\mn@doi@}
  {\mn@doi@[]}}
\def\mn@doi@[#1]#2{\def\@tempa{#1}\ifx\@tempa\@empty \href
  {http://dx.doi.org/#2} {doi:#2}\else \href {http://dx.doi.org/#2} {#1}\fi
  \endgroup}
\def\mn@eprint#1#2{\mn@eprint@#1:#2::\@nil}
\def\mn@eprint@arXiv#1{\href {http://arxiv.org/abs/#1} {{\tt arXiv:#1}}}
\def\mn@eprint@dblp#1{\href {http://dblp.uni-trier.de/rec/bibtex/#1.xml}
  {dblp:#1}}
\def\mn@eprint@#1:#2:#3:#4\@nil{\def\@tempa {#1}\def\@tempb {#2}\def\@tempc
  {#3}\ifx \@tempc \@empty \let \@tempc \@tempb \let \@tempb \@tempa \fi \ifx
  \@tempb \@empty \def\@tempb {arXiv}\fi \@ifundefined
  {mn@eprint@\@tempb}{\@tempb:\@tempc}{\expandafter \expandafter \csname
  mn@eprint@\@tempb\endcsname \expandafter{\@tempc}}}

\bibitem[\protect\citeauthoryear{Abplanalp, Gozem, Krylov, Shingledecker,
  Herbst  \& Kaiser}{Abplanalp et~al.}{2016}]{abplanalp_study_2016}
Abplanalp M.~J.,  Gozem S.,  Krylov A.~I.,  Shingledecker C.~N.,  Herbst E.,
  Kaiser R.~I.,  2016, \mn@doi [PNAS] {10.1073/pnas.1604426113}, 113, 7727

\bibitem[\protect\citeauthoryear{Bacmann, Taquet, Faure, Kahane  \&
  Ceccarelli}{Bacmann et~al.}{2012}]{bacmann_detection_2012}
Bacmann A.,  Taquet V.,  Faure A.,  Kahane C.,   Ceccarelli C.,  2012, \mn@doi
  [A\&A] {10.1051/0004-6361/201219207}, 541, L12

\bibitem[\protect\citeauthoryear{Balucani, Ceccarelli  \& Taquet}{Balucani
  et~al.}{2015}]{balucani_formation_2015}
Balucani N.,  Ceccarelli C.,   Taquet V.,  2015, \mn@doi [MNRAS]
  {10.1093/mnrasl/slv009}, 449, L16

\bibitem[\protect\citeauthoryear{Bennett \& Kaiser}{Bennett \&
  Kaiser}{2007a}]{bennett_formation_2007}
Bennett C.~J.,  Kaiser R.~I.,  2007a, \mn@doi [ApJ] {10.1086/513267}, 660, 1289

\bibitem[\protect\citeauthoryear{Bennett \& Kaiser}{Bennett \&
  Kaiser}{2007b}]{bennett_formation_2007-1}
Bennett C.~J.,  Kaiser R.~I.,  2007b, \mn@doi [ApJ] {10.1086/516745}, 661, 899

\bibitem[\protect\citeauthoryear{Bennett, Osamura, Lebar  \& Kaiser}{Bennett
  et~al.}{2005}]{bennett_laboratory_2005}
Bennett C.~J.,  Osamura Y.,  Lebar M.~D.,   Kaiser R.~I.,  2005, \mn@doi [ApJ]
  {10.1086/452618}, 634, 698

\bibitem[\protect\citeauthoryear{Berger, Coursey  \& Zucker}{Berger
  et~al.}{1999}]{berger_estar_1999}
Berger M.~J.,  Coursey J.~S.,   Zucker M.~A.,  1999

\bibitem[\protect\citeauthoryear{Bertin et~al.,}{Bertin
  et~al.}{2013}]{bertin_indirect_2013}
Bertin M.,  et~al., 2013, \mn@doi [ApJ] {10.1088/0004-637X/779/2/120}, 779, 120

\bibitem[\protect\citeauthoryear{Bethe}{Bethe}{1932}]{bethe_bremsformel_1932}
Bethe H.,  1932, \mn@doi [Z. Physik] {10.1007/BF01342532}, 76, 293

\bibitem[\protect\citeauthoryear{Blagojevic, Petrie  \& Bohme}{Blagojevic
  et~al.}{2003}]{blagojevic_gas-phase_2003}
Blagojevic V.,  Petrie S.,   Bohme D.~K.,  2003, \mn@doi [MNRAS]
  {10.1046/j.1365-8711.2003.06351.x}, 339, L7

\bibitem[\protect\citeauthoryear{Boyer, Rivas, Tran, Verish  \&
  Arumainayagam}{Boyer et~al.}{2016}]{boyer_role_2016}
Boyer M.~C.,  Rivas N.,  Tran A.~A.,  Verish C.~A.,   Arumainayagam C.~R.,
  2016, \mn@doi [Surface Science] {10.1016/j.susc.2016.03.012}, 652, 26

\bibitem[\protect\citeauthoryear{Brown, Crofts, Gardner, Godfrey, Robinson  \&
  Whiteoak}{Brown et~al.}{1975}]{brown_discovery_1975}
Brown R.~D.,  Crofts J.~G.,  Gardner F.~F.,  Godfrey P.~D.,  Robinson B.~J.,
  Whiteoak J.~B.,  1975, \mn@doi [ApJ] {10.1086/181769}, 197, L29

\bibitem[\protect\citeauthoryear{Carroll, Drouin  \& Weaver}{Carroll
  et~al.}{2010}]{carroll_submillimeter_2010}
Carroll P.~B.,  Drouin B.~J.,   Weaver S. L.~W.,  2010, \mn@doi [ApJ]
  {10.1088/0004-637X/723/1/845}, 723, 845

\bibitem[\protect\citeauthoryear{Cernicharo, Marcelino, Roueff, Gerin,
  Jiménez-Escobar  \& Muñoz~Caro}{Cernicharo
  et~al.}{2012}]{cernicharo_discovery_2012}
Cernicharo J.,  Marcelino N.,  Roueff E.,  Gerin M.,  Jiménez-Escobar A.,
  Muñoz~Caro G.~M.,  2012, \mn@doi [ApJ] {10.1088/2041-8205/759/2/L43}, 759,
  L43

\bibitem[\protect\citeauthoryear{Chang \& Herbst}{Chang \&
  Herbst}{2016}]{chang_unified_2016}
Chang Q.,  Herbst E.,  2016, \mn@doi [ApJ] {10.3847/0004-637X/819/2/145}, 819,
  145

\bibitem[\protect\citeauthoryear{Dalgarno, Griffing  \& Bates}{Dalgarno
  et~al.}{1958}]{dalgarno_energy_1958}
Dalgarno A.,  Griffing G.~W.,   Bates D.~R.,  1958, \mn@doi [Proc. of the R.
  Soc. of London. Series A. Math. and Phys. Sci.] {10.1098/rspa.1958.0253},
  248, 415

\bibitem[\protect\citeauthoryear{Dartois, Chabot, Barkach, Rothard, Augé,
  Agnihotri, Domaracka  \& Boduch}{Dartois
  et~al.}{2019}]{dartois_non-thermal_2019}
Dartois E.,  Chabot M.,  Barkach T.~I.,  Rothard H.,  Augé B.,  Agnihotri
  A.~N.,  Domaracka A.,   Boduch P.,  2019, \mn@doi [A\&A]
  {10.1051/0004-6361/201834787}, 627, A55

\bibitem[\protect\citeauthoryear{Ehrenfreund \& Charnley}{Ehrenfreund \&
  Charnley}{2000}]{ehrenfreund_organic_2000}
Ehrenfreund P.,  Charnley S.~B.,  2000, \mn@doi [ARA&A]
  {10.1146/annurev.astro.38.1.427}, 38, 427

\bibitem[\protect\citeauthoryear{Ellder et~al.,}{Ellder
  et~al.}{1980}]{ellder_methyl_1980}
Ellder J.,  et~al., 1980, \mn@doi [ApJ] {10.1086/183410}, 242, L93

\bibitem[\protect\citeauthoryear{Fedoseev, Cuppen, Ioppolo, Lamberts  \&
  Linnartz}{Fedoseev et~al.}{2015}]{fedoseev_experimental_2015}
Fedoseev G.,  Cuppen H.~M.,  Ioppolo S.,  Lamberts T.,   Linnartz H.,  2015,
  \mn@doi [MNRAS] {10.1093/mnras/stu2603}, 448, 1288

\bibitem[\protect\citeauthoryear{Fuchs, Cuppen, Ioppolo, Romanzin, Bisschop,
  Andersson, Dishoeck  \& Linnartz}{Fuchs
  et~al.}{2009}]{fuchs_hydrogenation_2009}
Fuchs G.~W.,  Cuppen H.~M.,  Ioppolo S.,  Romanzin C.,  Bisschop S.~E.,
  Andersson S.,  Dishoeck E. F.~v.,   Linnartz H.,  2009, \mn@doi [A\&A]
  {10.1051/0004-6361/200810784}, 505, 629

\bibitem[\protect\citeauthoryear{Garrod}{Garrod}{2013}]{garrod_three-phase_2013}
Garrod R.~T.,  2013, \mn@doi [ApJ] {10.1088/0004-637X/765/1/60}, 765, 60

\bibitem[\protect\citeauthoryear{Garrod \& Herbst}{Garrod \&
  Herbst}{2006}]{garrod_formation_2006}
Garrod R.~T.,  Herbst E.,  2006, \mn@doi [A\&A] {10.1051/0004-6361:20065560},
  457, 927

\bibitem[\protect\citeauthoryear{Garrod, Wakelam  \& Herbst}{Garrod
  et~al.}{2007}]{garrod_non-thermal_2007}
Garrod R.~T.,  Wakelam V.,   Herbst E.,  2007, \mn@doi [A\&A]
  {10.1051/0004-6361:20066704}, 467, 1103

\bibitem[\protect\citeauthoryear{Garrod, Weaver  \& Herbst}{Garrod
  et~al.}{2008}]{garrod_complex_2008}
Garrod R.~T.,  Weaver S. L.~W.,   Herbst E.,  2008, \mn@doi [ApJ]
  {10.1086/588035}, 682, 283

\bibitem[\protect\citeauthoryear{Graedel \& McGill}{Graedel \&
  McGill}{1982}]{graedel_graphical_1982}
Graedel T.~E.,  McGill R.,  1982, \mn@doi [Science]
  {10.1126/science.215.4537.1191}, 215, 1191

\bibitem[\protect\citeauthoryear{Hasegawa \& Herbst}{Hasegawa \&
  Herbst}{1993}]{hasegawa_new_1993}
Hasegawa T.~I.,  Herbst E.,  1993, \mn@doi [MNRAS] {10.1093/mnras/261.1.83},
  261, 83

\bibitem[\protect\citeauthoryear{Herbst \& Millar}{Herbst \&
  Millar}{2008}]{smith_chemistry_2008}
Herbst E.,  Millar T.~J.,  2008, in , Low {Temperatures} and {Cold}
  {Molecules}.
PUBLISHED BY IMPERIAL COLLEGE PRESS AND DISTRIBUTED BY WORLD SCIENTIFIC
  PUBLISHING CO., pp 1--54, \mn@doi{10.1142/9781848162105_0001}, \url
  {http://www.worldscientific.com/doi/abs/10.1142/9781848162105_0001}

\bibitem[\protect\citeauthoryear{Hollis, Lovas  \& Jewell}{Hollis
  et~al.}{2000}]{hollis_interstellar_2000}
Hollis J.~M.,  Lovas F.~J.,   Jewell P.~R.,  2000, \mn@doi [ApJ]
  {10.1086/312881}, 540, L107

\bibitem[\protect\citeauthoryear{Hudson, Moore  \& Cook}{Hudson
  et~al.}{2005}]{hudson_ir_2005}
Hudson R.~L.,  Moore M.~H.,   Cook A.~M.,  2005, \mn@doi [Advances in Space
  Research] {10.1016/j.asr.2005.01.017}, 36, 184

\bibitem[\protect\citeauthoryear{Huntress \& Mitchell}{Huntress \&
  Mitchell}{1979}]{huntress_synthesis_1979}
Huntress Jr. W.~T.,  Mitchell G.~F.,  1979, \mn@doi [ApJ] {10.1086/157207},
  231, 456

\bibitem[\protect\citeauthoryear{Jenkins}{Jenkins}{2009}]{jenkins_unified_2009}
Jenkins E.~B.,  2009, \mn@doi [ApJ] {10.1088/0004-637X/700/2/1299}, 700, 1299

\bibitem[\protect\citeauthoryear{Jiménez-Serra et~al.,}{Jiménez-Serra
  et~al.}{2016}]{jimenez-serra_spatial_2016}
Jiménez-Serra I.,  et~al., 2016, \mn@doi [ApJL] {10.3847/2041-8205/830/1/L6},
  830, L6

\bibitem[\protect\citeauthoryear{Jin \& Garrod}{Jin \&
  Garrod}{2020}]{Jin_Garrod_2020}
Jin M.,  Garrod R.~T.,  2020, ApJS, in press

\bibitem[\protect\citeauthoryear{Johnson}{Johnson}{1990}]{johnson_energetic_1990}
Johnson R.~E.,  1990, Energetic Charged-Particle Interactions with Atmospheres
  and Surfaces, X, 232 pp. 84 figs., 28 tabs.. Springer-Verlag Berlin
  Heidelberg New York. Also Physics and Chemistry in Space, volume 19

\bibitem[\protect\citeauthoryear{Kalvāns \& Kalnin}{Kalvāns \&
  Kalnin}{2019}]{kalvans_chemical_2019}
Kalvāns J.,  Kalnin J.~R.,  2019, \mn@doi [MNRAS] {10.1093/mnras/stz1010},
  486, 2050

\bibitem[\protect\citeauthoryear{Keller-Rudek, Moortgat, Sander  \&
  Sörensen}{Keller-Rudek et~al.}{2013}]{keller-rudek_mpi-mainz_2013}
Keller-Rudek H.,  Moortgat G.~K.,  Sander R.,   Sörensen R.,  2013, \mn@doi
  [Earth Syst. Sci. Data] {10.5194/essd-5-365-2013}, 5, 365

\bibitem[\protect\citeauthoryear{Laas, Garrod, Herbst  \& Widicus~Weaver}{Laas
  et~al.}{2011}]{laas_contributions_2011}
Laas J.~C.,  Garrod R.~T.,  Herbst E.,   Widicus~Weaver S.~L.,  2011, \mn@doi
  [ApJ] {10.1088/0004-637X/728/1/71}, 728, 71

\bibitem[\protect\citeauthoryear{Lias, Bartmess, Liebman, Holmes, Levin  \&
  Mallard}{Lias et~al.}{2018}]{lias_ion_2018}
Lias S.~G.,  Bartmess J.~E.,  Liebman J.~F.,  Holmes J.~L.,  Levin R.~D.,
  Mallard W.~G.,  2018, in , {NIST} {Chemistry} {WebBook}, {NIST} {Standard}
  {Reference} {Database} {Number} 69.
National Institute of Standards and Technology, Gaithersburg MD, 20899, \url
  {https://doi.org/10.18434/T4D303}

\bibitem[\protect\citeauthoryear{McGuire, Burkhardt, Kalenskii, Shingledecker,
  Remijan, Herbst  \& McCarthy}{McGuire et~al.}{2018}]{mcguire_detection_2018}
McGuire B.~A.,  Burkhardt A.~M.,  Kalenskii S.,  Shingledecker C.~N.,  Remijan
  A.~J.,  Herbst E.,   McCarthy M.~C.,  2018, \mn@doi [Science]
  {10.1126/science.aao4890}, 359, 202

\bibitem[\protect\citeauthoryear{Mehringer, Snyder, Miao  \& Lovas}{Mehringer
  et~al.}{1997}]{mehringer_detection_1997}
Mehringer D.~M.,  Snyder L.~E.,  Miao Y.,   Lovas F.~J.,  1997, \mn@doi [ApJ]
  {10.1086/310612}, 480, L71

\bibitem[\protect\citeauthoryear{Minissale, Dulieu, Cazaux  \& Hocuk}{Minissale
  et~al.}{2016}]{minissale_dust_2016}
Minissale M.,  Dulieu F.,  Cazaux S.,   Hocuk S.,  2016, \mn@doi [A\&A]
  {10.1051/0004-6361/201525981}, 585, A24

\bibitem[\protect\citeauthoryear{Neufeld, Wolfire  \& Schilke}{Neufeld
  et~al.}{2005}]{neufeld_chemistry_2005}
Neufeld D.~A.,  Wolfire M.~G.,   Schilke P.,  2005, \mn@doi [ApJ]
  {10.1086/430663}, 628, 260

\bibitem[\protect\citeauthoryear{Remijan, Snyder, Liu, Mehringer  \&
  Kuan}{Remijan et~al.}{2002}]{remijan_acetic_2002}
Remijan A.,  Snyder L.~E.,  Liu S.-Y.,  Mehringer D.,   Kuan Y.-J.,  2002,
  \mn@doi [ApJ] {10.1086/341627}, 576, 264

\bibitem[\protect\citeauthoryear{Remijan, Snyder, Friedel, Liu  \&
  Shah}{Remijan et~al.}{2003}]{remijan_survey_2003}
Remijan A.,  Snyder L.~E.,  Friedel D.~N.,  Liu S.-Y.,   Shah R.~Y.,  2003,
  \mn@doi [ApJ] {10.1086/374890}, 590, 314

\bibitem[\protect\citeauthoryear{Remijan, Shiao, Friedel, Meier  \&
  Snyder}{Remijan et~al.}{2004}]{remijan_survey_2004}
Remijan A.,  Shiao Y.-S.,  Friedel D.~N.,  Meier D.~S.,   Snyder L.~E.,  2004,
  \mn@doi [ApJ] {10.1086/425266}, 617, 384

\bibitem[\protect\citeauthoryear{Rothard, Domaracka, Boduch, Palumbo,
  Strazzulla, Silveira  \& Dartois}{Rothard
  et~al.}{2017}]{rothard_modification_2017}
Rothard H.,  Domaracka A.,  Boduch P.,  Palumbo M.~E.,  Strazzulla G.,
  Silveira E. F.~d.,   Dartois E.,  2017, \mn@doi [J. Phys. B: At. Mol. Opt.
  Phys.] {10.1088/1361-6455/50/6/062001}, 50, 062001

\bibitem[\protect\citeauthoryear{Ruaud, Loison, Hickson, Gratier, Hersant  \&
  Wakelam}{Ruaud et~al.}{2015}]{ruaud_modelling_2015}
Ruaud M.,  Loison J.~C.,  Hickson K.~M.,  Gratier P.,  Hersant F.,   Wakelam
  V.,  2015, \mn@doi [Mon Not R Astron Soc] {10.1093/mnras/stu2709}, 447, 4004

\bibitem[\protect\citeauthoryear{Ruaud, Wakelam  \& Hersant}{Ruaud
  et~al.}{2016}]{ruaud_gas_2016}
Ruaud M.,  Wakelam V.,   Hersant F.,  2016, \mn@doi [MNRAS]
  {10.1093/mnras/stw887}, 459, 3756

\bibitem[\protect\citeauthoryear{Shingledecker \& Herbst}{Shingledecker \&
  Herbst}{2018}]{shingledecker_general_2018}
Shingledecker C.~N.,  Herbst E.,  2018, \mn@doi [Phys. Chem. Chem. Phys.]
  {10.1039/C7CP05901A}, 20, 5359

\bibitem[\protect\citeauthoryear{Shingledecker, Tennis, Gal  \&
  Herbst}{Shingledecker et~al.}{2018}]{shingledecker_cosmic-ray-driven_2018}
Shingledecker C.~N.,  Tennis J.,  Gal R.~L.,   Herbst E.,  2018, \mn@doi [ApJ]
  {10.3847/1538-4357/aac5ee}, 861, 20

\bibitem[\protect\citeauthoryear{Shingledecker, Vasyunin, Herbst  \&
  Caselli}{Shingledecker et~al.}{2019a}]{shingledecker_simulating_2019}
Shingledecker C.~N.,  Vasyunin A.,  Herbst E.,   Caselli P.,  2019a, \mn@doi
  [ApJ] {10.3847/1538-4357/ab16d5}, 876, 140

\bibitem[\protect\citeauthoryear{Shingledecker, Álvarez Barcia, Korn  \&
  Kästner}{Shingledecker et~al.}{2019b}]{shingledecker_case_2019}
Shingledecker C.~N.,  Álvarez Barcia S.,  Korn V.~H.,   Kästner J.,  2019b,
  \mn@doi [ApJ] {10.3847/1538-4357/ab1d4a}, 878, 80

\bibitem[\protect\citeauthoryear{Shingledecker, Lamberts, Laas, Vasyunin,
  Herbst, Kästner  \& Caselli}{Shingledecker
  et~al.}{2020a}]{shingledecker_efficient_2020}
Shingledecker C.~N.,  Lamberts T.,  Laas J.~C.,  Vasyunin A.,  Herbst E.,
  Kästner J.,   Caselli P.,  2020a, \mn@doi [ApJ] {10.3847/1538-4357/ab5360},
  888, 52

\bibitem[\protect\citeauthoryear{Shingledecker, Molpeceres, Rivilla, Majumdar
  \& Kästner}{Shingledecker et~al.}{2020b}]{shingledecker_isomers_2020}
Shingledecker C.~N.,  Molpeceres G.,  Rivilla V.~M.,  Majumdar L.,   Kästner
  J.,  2020b, ApJ, Accepted

\bibitem[\protect\citeauthoryear{Skouteris, Balucani, Ceccarelli, Vazart,
  Puzzarini, Barone, Codella  \& Lefloch}{Skouteris
  et~al.}{2018}]{skouteris_genealogical_2018}
Skouteris D.,  Balucani N.,  Ceccarelli C.,  Vazart F.,  Puzzarini C.,  Barone
  V.,  Codella C.,   Lefloch B.,  2018, \mn@doi [ApJ]
  {10.3847/1538-4357/aaa41e}, 854, 135

\bibitem[\protect\citeauthoryear{Soma, Sakai, Watanabe  \& Yamamoto}{Soma
  et~al.}{2018}]{soma_complex_2018}
Soma T.,  Sakai N.,  Watanabe Y.,   Yamamoto S.,  2018, \mn@doi [ApJ]
  {10.3847/1538-4357/aaa70c}, 854, 116

\bibitem[\protect\citeauthoryear{Spitzer \& Tomasko}{Spitzer \&
  Tomasko}{1968}]{spitzer_heating_1968}
Spitzer Jr. L.,  Tomasko M.~G.,  1968, \mn@doi [ApJ] {10.1086/149610}, 152, 971

\bibitem[\protect\citeauthoryear{Taquet, Wirström, Charnley, Faure,
  López-Sepulcre  \& Persson}{Taquet et~al.}{2017}]{taquet_chemical_2017}
Taquet V.,  Wirström E.~S.,  Charnley S.~B.,  Faure A.,  López-Sepulcre A.,
  Persson C.~M.,  2017, \mn@doi [A\&A] {10.1051/0004-6361/201630023}, 607, A20

\bibitem[\protect\citeauthoryear{Vastel, Ceccarelli, Lefloch  \&
  Bachiller}{Vastel et~al.}{2014}]{vastel_origin_2014}
Vastel C.,  Ceccarelli C.,  Lefloch B.,   Bachiller R.,  2014, \mn@doi [ApJ]
  {10.1088/2041-8205/795/1/L2}, 795, L2

\bibitem[\protect\citeauthoryear{Vasyunin \& Herbst}{Vasyunin \&
  Herbst}{2012}]{vasyunin_unified_2012}
Vasyunin A.~I.,  Herbst E.,  2012, \mn@doi [ApJ] {10.1088/0004-637X/762/2/86},
  762, 86

\bibitem[\protect\citeauthoryear{Vasyunin \& Herbst}{Vasyunin \&
  Herbst}{2013}]{vasyunin_unified_2013}
Vasyunin A.~I.,  Herbst E.,  2013, \mn@doi [ApJ] {10.1088/0004-637X/762/2/86},
  762, 86

\bibitem[\protect\citeauthoryear{Wakelam \& Herbst}{Wakelam \&
  Herbst}{2008}]{wakelam_polycyclic_2008}
Wakelam V.,  Herbst E.,  2008, \mn@doi [ApJ] {10.1086/587734}, 680, 371

\bibitem[\protect\citeauthoryear{Wakelam et~al.,}{Wakelam
  et~al.}{2012}]{wakelam_kinetic_2012}
Wakelam V.,  et~al., 2012, \mn@doi [ApJS] {10.1088/0067-0049/199/1/21}, 199, 21

\bibitem[\protect\citeauthoryear{Woods, Kelly, Viti, Slater, Brown, Puletti,
  Burke  \& Raza}{Woods et~al.}{2012}]{woods_formation_2012}
Woods P.~M.,  Kelly G.,  Viti S.,  Slater B.,  Brown W.~A.,  Puletti F.,  Burke
  D.~J.,   Raza Z.,  2012, \mn@doi [ApJ] {10.1088/0004-637X/750/1/19}, 750, 19

\bibitem[\protect\citeauthoryear{Woods, Slater, Raza, Viti, Brown  \&
  Burke}{Woods et~al.}{2013}]{woods_glycolaldehyde_2013}
Woods P.~M.,  Slater B.,  Raza Z.,  Viti S.,  Brown W.~A.,   Burke D.~J.,
  2013, \mn@doi [ApJ] {10.1088/0004-637X/777/2/90}, 777, 90

\bibitem[\protect\citeauthoryear{Xue, Remijan, Brogan, Hunter, Herbst  \&
  McGuire}{Xue et~al.}{2019a}]{xue_alma_2019-1}
Xue C.,  Remijan A.~J.,  Brogan C.~L.,  Hunter T.~R.,  Herbst E.,   McGuire
  B.~A.,  2019a, arXiv e-prints, p. arXiv:1907.07117

\bibitem[\protect\citeauthoryear{Xue, Remijan, Burkhardt  \& Herbst}{Xue
  et~al.}{2019b}]{xue_alma_2019}
Xue C.,  Remijan A.~J.,  Burkhardt A.~M.,   Herbst E.,  2019b, \mn@doi [ApJ]
  {10.3847/1538-4357/aaf738}, 871, 112

\bibitem[\protect\citeauthoryear{Öberg, van Dishoeck  \& Linnartz}{Öberg
  et~al.}{2009}]{oberg_photodesorption_2009}
Öberg K.~I.,  van Dishoeck E.~F.,   Linnartz H.,  2009, \mn@doi [A\&A]
  {10.1051/0004-6361/200810207}, 496, 281

\makeatother
\end{thebibliography}



\section*{Supporting Information}
\label{sec:Support_info}
Supplementary Tables are available online through MNRAS.
Table S1 is a full list of thermal reactions added to the base KIDA network, both gas phase and solid phase.
Table S2 lists new radiolysis reactions not previously published, and Table S3 lists new suprathermal reactions with reactions both from this paper and previously studied reactions.


\bsp	
\label{lastpage}
\end{document}